\documentclass[fleqn,usenatbib]{mnras}

\usepackage[T1]{fontenc}

\DeclareRobustCommand{\VAN}[3]{#2}
\let\VANthebibliography\thebibliography
\def\thebibliography{\DeclareRobustCommand{\VAN}[3]{##3}\VANthebibliography}

\usepackage{graphicx}	
\usepackage{amsmath}	
\usepackage{amssymb}	
\usepackage{newtxtext,newtxmath}

\title[Hierarchical modelling of Spitzer data]{Atmospheric characterization of hot Jupiters using hierarchical models of \textit{Spitzer} observations}

\author[D. Keating and N. Cowan]{
Dylan Keating,$^{1}$\thanks{E-mail: dylan.keating@mail.mcgill.ca}
Nicolas B. Cowan,$^{1,2}$
\\
$^{1}$Department of Physics, McGill University, Montr\'{e}al, QC H3A 2T8, Canada\\
$^{2}$Department of Earth \& Planetary Sciences, McGill University, Montr\'{e}al, QC H3A 2T8, Canada
}

\date{Submitted 2021 February 25}

\pubyear{2021}

\begin{document}
\label{firstpage}
\pagerange{\pageref{firstpage}--\pageref{lastpage}}
\maketitle

\begin{abstract}
The field of exoplanet atmospheric characterization is trending towards comparative studies involving many planetary systems, and using Bayesian hierarchical modelling is a natural next step. Here we demonstrate two use cases.  We first use hierarchical modelling to quantify variability in repeated observations by reanalyzing a suite of ten \textit{Spitzer} secondary eclipse observations of the hot Jupiter XO-3b. We compare three models: one where we fit ten separate eclipse depths, one where we use a single eclipse depth for all ten observations, and a hierarchical model. By comparing the Widely Applicable Information Criterion of each model, we show that the hierarchical model is preferred over the others. The hierarchical model yields less scatter across the suite of eclipse depths---and higher precision on the individual eclipse depths---than does fitting the observations separately. We find that the hierarchical eclipse depth uncertainty is larger than the uncertainties on the individual eclipse depths, which suggests either slight astrophysical variability or that single eclipse observations underestimate the true eclipse depth uncertainty. Finally, we fit a suite of published dayside brightness measurements for 37 planets using a hierarchical model of brightness temperature vs irradiation temperature. The hierarchical model gives tighter constraints on the individual brightness temperatures than the non-hierarchical model. Although we tested hierarchical modelling on \textit{Spitzer} eclipse data of hot Jupiters, it is applicable to observations of smaller planets like hot neptunes and super earths, as well as for photometric and spectroscopic transit or phase curve observations.
\end{abstract}

\begin{keywords}
planets and satellites: individual (XO-3b) -- techniques: photometric
\end{keywords}



\section{Introduction} \label{sec:intro}
Although the \textit{Spitzer Space Telescope} wasn't designed for exoplanet science, it was a workhorse for the field \citep[for a recent review, read][]{2020NatAs...4..453D}. In particular, observations of exoplanet transits, secondary eclipses, and phase curves with \emph{Spitzer's} Infrared Array Camera have been used to characterize the atmospheres of over a hundred transiting planets.

Substantial progress has been made towards a statistical understanding of exoplanetary atmospheres \citep{Cowan2011,Sing2016,Schwartz2015,Schwartz2017,Parmentier2018,Zhang2018,Keating2019,Keating2020,Baxter2020,Bell2021}. Many of the planets in these studies had been analyzed using disparate reduction and analysis pipelines, but researchers have started uniformly analyzing observations of multiple planets using a single pipeline. \citet{Garhart2020} independently reduced and analyzed 78 eclipse depths from 36 planets and found that hotter planets had higher brightness temperatures at 4.5~$\mu$m than at 3.6~$\mu$m. \citet{Bell2021} reanalyzed every available \textit{Spitzer} 4.5~$\mu$m hot Jupiter phase curve using an open-source reduction and analysis pipeline, confirming several previously reported trends.  

In this work we outline a complementary way to further the statistical understanding of exoplanet atmospheres: fitting measurements from multiple planets simultaneously using hierarchical models to robustly infer trends.

\subsection{Spitzer Systematics}
Exoplanet observations taken with \textit{Spitzer}'s Infrared Array Camera \citep[IRAC;][]{Fazio2004} are dominated by systematics noise. The systematics are driven by intrapixel sensitivity variations on the detector and by now are well characterized  \citep{Ingalls2016}. Detector systematics are typically fitted simultaneously with the astrophysical signal of interest. Each transit, secondary eclipse, and phase curve yields information about the IRAC detector sensitivity, but typically this information is not shared between observations. 

Since the \textit{Spitzer} systematics are a function of the centroid location on the pixel, efforts have been made to map the detector sensitivity independently using observations of quiet stars \citep{Ingalls2012,Krick2020,May2020}. The flux of a calibration star should be constant as a function of time, so any deviation must be due to the centroid moving across the detector as the telescope pointing drifts. A crucial assumption for this approach is that the \textit{Spitzer} systematics do not vary with time, and that they are not dependent on the brightness of the star. \citet{Ingalls2012} and \citet{May2020} approached the problem by explicitly calculating the detector sensitivity, while \citet{Krick2020} used a machine learning technique called random forests to look for patterns in the systematics.

Other approaches do not assume anything explicit about the detector sensitivity. Independent component analysis \citep{Waldmann2012,Morello2014,Morello2016} separates the signal into additive subcomponents using blind source separation, with the idea being that one of these signals is the astrophysical signal. In another approach, \citet{Morvan2020} used the baseline signal before and after a transit to learn and predict the in transit detector systematics using a machine learning technique known as Long short-term memory networks. 

In this work, we opted to parameterize and fit the detector systematics simultaneously with the astrophysical signal to account for any correlations between the two.

\subsection{Hierarchical Models}
Bayesian hierarchical models \citep{BDA} are routinely used in other fields because they offer a natural way to infer higher level trends in a dataset and can increase measurement precision. They are gaining traction in exoplanet studies: for example, to study the mass-radius \citep{Teske2020} and mass-radius-period \citep{Neil2020} relations, and radius inflation of hot Jupiters \citep{Sarkis2020,Thorngren2021}. Hierarchical models have not yet been applied to atmospheric characterization of exoplanets.

There is one major difference between a typical Bayesian model and a hierarchical one. In a traditional Bayesian model, we estimate the probability distribution of our model parameters given our observed data and the prior probability of each model parameter. The prior distribution encodes our previous knowledge about the most likely values of the parameters and is specified before fitting the model. In a hierarchical model, however, the prior distributions themselves are parameterized using so-called hyperparameters. The hyperparameters become part of the model and are fitted simultaneously with the other parameters of interest. As we explain in the next section, this naturally represents how our intuition pools information across observations. It also helps to tame models by compromising between overfitting and underfitting.  

Hierarchical models should be used whenever the data allow us to refine our knowledge of the prior distribution, which happens when a certain quantity is measured multiple times. A natural example in exoplanet science is repeated \textit{Spitzer} observations of the same target. To demonstrate, we start with the archetypical suite of of ten \textit{Spitzer} IRAC Channel 2 (4.5~$\mu$m) secondary eclipses of the eccentric hot Jupiter XO-3b \citep{Wong2014}. Below we explain the model and present our results. Afterwards, we show how we extend the model to fit multiple eclipses from different planets simultaneously and present results from fitting the eclipse data from \citet{Garhart2020} with a hierarchical model.

\section{Hierarchical Model of XO-3b Eclipses}

In the \textit{Spitzer} data challenge, several groups analyzed ten secondary eclipses of XO-3b in order to test the repeatability and accuracy of various decorrelation techniques  \citep{Ingalls2016}. The reduced archival data from the data challenge are publicly available, so we downloaded them rather than reducing them ourselves. 

For XO-3b and other planets with repeated secondary eclipse observations, the eclipses have usually been fitted separately from one another, with a separate eclipse depth parameter for each observation \citep{Ingalls2016,Kilpatrick2020}. In other cases, a single eclipse depth parameter has been used to simultaneously fit multiple secondary eclipse measurements \citep{Wong2014}. This is also what is typically done for phase curves that are bracketed by two eclipses \citep{Cowan2012, Bell2021}.   

However, neither approach quite matches what our intuition tells us. Because we are measuring the same thing each time, fitting the eclipse observations separately amounts to overfitting the individual observations, and fitting a single eclipse parameter amounts to underfitting all of the observations. 
If we observe one secondary eclipse, we would expect that the next one we observe would have a similar--- but not identical--- depth, due to measurement uncertainty, if not astrophysical variability. The second eclipse we observe would also change our beliefs about the first one. Each measurement of the planet's eclipse depth can be thought of as a draw from a distribution, with some variance. With enough measurements, the shape of this distribution can be inferred. A hierarchical model naturally takes all of this into account by fitting for the parameters that describe the higher level distribution simultaneously with the astrophysical signal of each observation.

Bayesian analysis requires us to specify priors on the parameters we are trying to infer. We can write down our prior on the $i$th eclipse depth as
\begin{equation}
    D_{i} \sim \mathcal{N}(\mu, \sigma),
\end{equation}
where we have used the tilde shorthand $(\sim)$ to mean that the eclipse depth is drawn from a normal distribution centered on $\mu$ with a standard deviation of $\sigma$; $\mu$ and $\sigma$ are hyperparameters. In a non-hierarchical model, we would specify $\mu$ and $\sigma$ to represent our prior expectations of what $D_{i}$ could be. After fitting, we would get a separate posterior distribution for each eclipse depth. 

In a hierarchical model, we instead make $\mu$ and $\sigma$ parameters and fit them simultaneously with the ten eclipse depths. We represent our beliefs about hyperparameters $\mu$ and $\sigma$ with hyperpriors. This allows each eclipse observation to inform the others, by pulling the eclipse depths closer to the mode of the distribution of $\mu$. This is known as Bayesian shrinkage. After fitting, we get a posterior distribution for each eclipse depth, as well as for $\mu$ and $\sigma$. 

In the limit that $\sigma$ goes to infinity, the hierarchical model is equivalent to the model with completely separate eclipse depths. Likewise when $\sigma$ goes to zero, it is equivalent to the single eclipse depth model. A hierarchical model empirically fits for the amount of pooling based on what is most consistent with the observations.

\subsection{Priors}
Priors are necessary in a fit to encode prior knowledge, as well as to properly sample a model. In all cases, we use weakly informative priors rather than flat, ``uninformative'' priors. A flat prior is equivalent to saying that all values of eclipse depth are equally likely, even extremely large, unphysical values. Instead, we chose to place a normal prior with a large standard deviation so that we kept the predicted values within the right order of magnitude. Half-normal priors or wide normal priors are unlikely to introduce much bias into the parameter estimates and can make sampling more efficient. Flat priors are discouraged in practice because we usually have at least some vague knowledge of the range of values a parameter can take \citep{Gelman2017}.

\subsection{Astrophysical Model}
The astrophysical model for each observation was a secondary eclipse. We used STARRY \citep{Luger2019} to compute the shape of each eclipse, with the depth and time of eclipse left as free parameters. We fixed the radius of the planet and host star, the orbital period, ratio of semi-major axis to stellar radius, orbital inclination, longitude of periastron and eccentricity to the literature values. 

To get a rough upper limit on the eclipse depth, we used the parameterization of \citet{Cowan2011} to calculate the maximum dayside temperature, in the limit of a Bond albedo of zero and no heat recirculation:
\begin{equation}
    T_{\rm d, max} = T_{\rm eff}\sqrt{\frac{R_{\star}}{a}}\left(\frac{2}{3}\right)^{1/4} \label{TdayMax}.
\end{equation} Here $T_{\rm eff}$ is the stellar effective temperature, and $a/R_{\star}$ is the ratio of semimajor axis to stellar radius. We note that this equation assumes a circular orbit, while XO-3b is on an eccentric orbit \citep[$e=0.28$;][]{Bonomo2017}. Nonetheless, it allows us to get an order of magnitude estimate of the eclipse depth.

The above temperature can be converted to an eclipse depth using
\begin{equation}
    D = \frac{B(\lambda, T_{\rm d})}{B(\lambda, T_{\star,4.5\mu m})} \left(\frac{R_{\rm p}}{R_{\star}}\right)^{2} \label{ED}
\end{equation}
where $B$ is the Planck function, and $T_{\star,4.5\mu m}$ is the brightness temperature of the star at 4.5~$\mu$m, which we calculated by integrating PHOENIX models \citep{Allard2011} over the \textit{Spitzer} bandpass \citep{Baxter2020}. We represent the eclipse depth when $T_{\rm d} = T_{\rm d,max}$ by $D_{max}$.

For the non-hierarchical model, we placed a wide prior on the eclipse depth to prevent biasing the value: $D \sim \mathcal{N}(D_{max}/2, D_{max}/2).$ For the time of eclipse, we let $\tau \sim \mathcal{N}(\Delta t/2,\Delta t/2)$ where $\Delta t$ is the duration of the observation, and time is measured from the start of the observation. We experimented with various priors and found that our resulting fits were consistent and not strongly dependent on the choice of priors.

For the hierarchical model, we used a wide Normal prior for the hierarchical mean: $\mu \sim \mathcal{N}(D_{max}/2, D_{max}/2).$ For the hierarchical standard deviation we used a weakly informative Half-Normal prior: $\sigma \sim$ Half-$\mathcal{N}(300\rm ppm).$ We then let the individual eclipse depths be drawn from the following higher level distribution: $D_{i} \sim \mathcal{N}(\mu,\sigma)$.

\subsection{Detector Systematics: Gaussian Processes}
The IRAC detector sensitivity in Channels 1 and 2 depends on the target centroid position on the detector. To parameterize this behaviour, we used a Gaussian process. The advantage of using a Gaussian process is that it doesn't require calculating the detector sensitivity explicitly, in contrast with polynomial models \citep{Cowan2012} or BLISS \citep{Stevenson2012a}.

When using Gaussian processes, we make the usual assumption that the data are normally distributed, but allow for covariance between data points. The likelihood function can be written 
\begin{equation}
    p(\rm data \vert \gamma) \sim \mathcal{N}(\mu_{\rm GP},\Sigma),
\end{equation}
where $\gamma$ represents the model parameters and independent variables, and $\mu_{\rm GP}$ is the mean function around which the data are distributed. The covariance function, $\Sigma$, is an $n\times n$ matrix where $n$ is the number of data. The entries along the diagonal of $\Sigma$ are the measurement uncertainties on each datum, which we denote by $\sigma_{\rm phot}$, and the off-diagonal entries are the covariance between data. When the off-diagonal elements are equal to zero, the likelihood function reduces to the usual assumption of independent Gaussian uncertainties.  

Although it is computationally intractable to fit each off-diagonal entry of the covariance matrix, they can be parameterized using a kernel function with a handful of parameters. We used the squared exponential kernel employed by \citet{Evans2015}
\begin{equation}
    \Sigma_{i j}=A \exp \left[-\left(\frac{x_{i}-x_{j}}{l_{x}}\right)^{2}-\left(\frac{y_{i}-y_{j}}{l_{y}}\right)^{2}\right],
\end{equation}
where $x$ and $y$ represent the centroid locations on the IRAC detector, in pixel coordinates. The terms $l_{x}$ and $l_{y}$ are the covariance lengthscales, and $A$ is the Gaussian process amplitude. The squared exponential kernel has the intuitive property that locations on the detector pixel that are close together should have similar sensitivity. If the length scales are fixed by the user rather than fitted for, this boils down to the Gaussian kernel regression of \cite{Ballard2010}, \citet{Knutson2012}, and \citet{Lewis2013}.

With no loss of generality, we can let the mean function be zero, and instead fit the residuals between the astrophysical model and observations which gives 
\begin{equation}
    p(\rm residuals \vert \gamma) \sim \mathcal{N}(0,\Sigma).
\end{equation}

We placed weakly informative inverse gamma priors on the lengthscales. We chose the parameters such that 99\% of the prior probability was between lengthscales 0 and 1, measured in pixels. This gives $p(l_{x}),p(l_{y}) \sim$ InverseGamma($\alpha$=11, $\beta$=5).

For the amplitude $A$, we used a weakly informative half normal prior: $A \sim$ Half-$\mathcal{N}(0,\Delta F/3)$, where $\Delta F$ is the range of observed flux values. By using a half normal prior, we weakly constrain the scale of the Gaussian process amplitude without introducing bias. We placed the same prior on the white noise uncertainty $\sigma_{\rm phot}$.

\subsection{The posterior}
For each eclipse, we fit the eclipse depth D, time of eclipse $\tau$, photometric uncertainty $\sigma_{\rm phot}$, Gaussian process lengthscales $l_{x}$ and $l_{y}$, and Gaussian process amplitude $A$. The planet-to-star flux ratio as a function of time $t$ is given by $F$. The $x$ and $y$ centroid locations are included as covariates. We also fit for the hierarchical eclipse depth mean $\mu$, and standard deviation $\sigma$.

The likelihood function for one eclipse is given by
\begin{equation}
    p(F \vert t, x, y, \sigma_{\rm phot}, D, \mu,\sigma, \tau , l_{x},l_{y})
\end{equation}
and the prior is
\begin{equation}
p(t, x, y, \sigma_{\rm phot}, D, \mu,\sigma,\tau , l_{x},l_{y}).
\end{equation} We form the posterior function by multiplying the likelihood and prior together:

\begin{equation}
    p(D,\mu,\sigma,\theta \vert F) \propto p(F \vert D,\mu,\sigma,\theta) p(D,\mu,\sigma,\theta)
\end{equation}
where we have used $\theta = \{t,x,y,\sigma_{\rm phot},\tau,l_{x},l_{y}\}$ for simplicity. 

First, note that the likelihood function does not depend on the hierarchical parameters directly, so we can remove $\mu$ and $\sigma$ from the brackets of the likelihood function. Second, the eclipse depth depends on the hierarchical parameters in this way:
\begin{equation}
    p(D \vert \mu,\sigma) = \frac{p(D, \mu,\sigma)}{p(\mu,\sigma)},
\end{equation}
where we have made use of Bayes' theorem.

This means we can rewrite the likelihood and prior to obtain:
\begin{equation}
    p(D,\mu,\sigma,\theta \vert F) \propto p(F \vert D,\theta)p(D\vert \mu,\sigma)p(\mu,\sigma,\theta).
\end{equation}It is this refactoring that makes a hierarchical model different from a non-hierarchical one.

To form the posterior of the full hierarchical model, we multiply the individual eclipse posteriors together:
\begin{equation}
\begin{split}
&\prod\limits_{i=1}^{n} p(D_{i},\mu,\sigma,\theta_{i} \vert F_{i}) \propto
\\
&(p(\mu,\sigma))^{n} \prod\limits_{i=1}^{n}p(F_{i} \vert D_{i},\theta_{i})p(D_{i}\vert \mu,\sigma)p(\theta_{i})
\end{split}
\end{equation}
where $n$ is the number of eclipse observations, 10 in the case of the XO-3b dataset. We have also used the fact that the priors on $\theta_{i}$ are independent from the priors on the hyperparameters $\mu$ and $\sigma$ to perform the separation $p(\mu,\sigma,\theta_{i}) = p(\mu,\sigma)p(\theta_i)$.

\subsection{Hamiltonian Monte Carlo}
We used the probabilistic programming package PyMC3 to build and sample from the model. PyMC3 uses Hamiltonian Monte Carlo (HMC), the state-of-the-art Markov Chain Monte Carlo (MCMC) algorithm,  to perform the sampling. HMC is more efficient than other MCMC algorithms, meaning it can effectively describe the posterior using fewer samples than other MCMC algorithms. For high dimensional models, and especially for high dimensional hierarchical models with pathological parameter spaces, the reduction in sample size afforded by HMC is all but necessary \citep{Betancourt2013}. 

Hamiltonian Monte Carlo works by treating probabilistic systems as if they are instead physical systems \citep{Betancourt2017}. The chains in an MCMC sampler move through parameter space to estimate the shape of the posterior; an equivalent physical system is the motion of a satellite orbiting a giant planet where the planet represents the mode of the probability distribution. The crucial step in HMC is to transform these trajectories from parameter space to momentum space (i.e., the space of the derivatives of the coordinates) using the Hamiltonian of the system, and sample from that instead by proposing a move in momentum space. By using conservation of energy, the HMC chains tend to remain in regions of high probability. In this way they efficiently traverse the typical set of the distribution, roughly defined as where the most of the probability mass of the posterior is concentrated.

To use Hamiltonian Monte Carlo, the gradient of the likelihood function, with respect to the parameters, is needed in order to construct the Hamiltonian of the system. PyMC3 does this using Theano, which is a deep learning library that allows for efficient manipulation of matrices \citep{Theano}. STARRY is also built on Theano and allows for analytic expressions and gradients in the general case of eclipse and phase mapping.  In our case, because we are dealing with eclipse-only observations and are able to neglect planetary limb darkening, the eclipse expressions reduce to the analytic expressions of \citet{MandelAgol2002}. Because astrophysical parameters tend to be correlated, we used the dense mass matrix HMC step from the \emph{exoplanet} package \citep{exoplanet}.

The biggest advantage is that HMC can diagnose problematic posteriors or models. Posteriors with pathological regions, such as high curvature, are hard for typical MCMC samplers to explore efficiently; hierarchical models exhibit such pathologies. This can lead to biases in the final results that are hard to diagnose because typical samplers do not have the ability to properly detect and respond to parameter spaces with extreme geometries. When an HMC chain gets stuck in a region of high curvature or otherwise behaves badly, it will diverge to infinity and the sampler keeps track of where this occurred. Divergences can often be eliminated by changing the acceptance probability of the sampler, or by reparameterizing the model.

\subsection{Model comparison: Information Criteria}
It is common among exoplanet scientists to use the Bayesian Information Criterion (BIC) or Aikake Information Criterion (AIC) to perform model comparison and selection \citep{Schwarz1978,Aikake1974}. Both criteria use the maximum likelihood and a complexity term to penalize overly complex models. These two information criteria describe slightly different things--- the AIC measures the relative predictive loss of a set of models, and the BIC measures how close each model is to the true model. In practice (at least in exoplanet science), they typically yield similar conclusions.  

A shortcoming shared by BIC and AIC is that they are accurate only when using flat priors, which are not recommended for most models \citep{Gelman2017}. The AIC also assumes that the posterior distributions are multivariate Gaussians. The priors are never flat for hierarchical models, which means we cannot use the AIC or BIC for model comparison. 

A more general model comparison tool is the Widely Applicable Information Criterion \citep[WAIC;][]{Watanabe2010}. The WAIC is Bayesian, uses the full fit posterior and, critically, makes no assumptions about the shape of the posterior or priors. The WAIC is easily computed from the full fit posterior \citep{StatisticalRethinking}: 
\begin{equation}
\mathrm{WAIC}(\rm data, \Theta)=-2\left(\operatorname{lppd}-\sum_{i} \operatorname{var}_{\theta} \left(\log p\left(F_{i} \mid D_{i}, \theta_{i}\right)\right) \right)
\end{equation}
where $\Theta$ is the posterior, $F_{i}$ stands for the i$^{th}$ eclipse observation, data refers to the entire suite of observations, and lppd stands for the log-pointwise predictive-density, \begin{equation}
    \operatorname{lppd} = \sum_{i} \log \left(\frac{1}{S} \sum_{s=0}^{S} p\left(F_{i} \mid D_{i}, \theta_{i,s}\right) \right),
\end{equation} where $S$ is the number of samples and $\theta_{i,s}$ is the $S^{\rm th}$ set of parameters for the $i^{\rm th}$ observation. The log predictive pointwise density is an estimate of how well the model would fit new, unseen data. The second term in the WAIC expression is a penalty term that penalizes overly complex models.

Once MCMC sampling has finished, the WAIC can be computed in a few lines of code using the MCMC chains. It is also possible to compute the standard error of the WAIC, something that is not possible with the BIC or AIC. If the difference in WAIC between two models is significantly larger than the standard error of the difference, then the model with the smaller WAIC is favoured over the other. If the difference in WAIC is smaller than the standard error of the difference, then the models make equally good predictions and there is no evidence to favour one over the other. 

Since all modern secondary eclipse, transit and phase curve analyses use Markov Chain Monte Carlo to sample and store the posterior draws, the WAIC is a better choice than BIC or AIC for model comparison. 

\subsection{Pooling the GP parameters}
We hypothesized that fitting a common set of Gaussian process amplitude and length scales across the suite of eclipses would yield more precise eclipse depths by sharing information about the detector sensitivity across the observations. Because we used Hamiltonian Monte Carlo, it was feasible to fully marginalize over the Gaussian process hyperparameters. However, we found that the fitted eclipse depths had nearly identical means and standard deviations between the shared and non-shared detector models of XO-3b. In practice we adopted the shared GP model for XO-3b because it has fewer parameters and is therefore easier to sample.
\subsection{Results}
\begin{figure}
\includegraphics[width=\columnwidth]{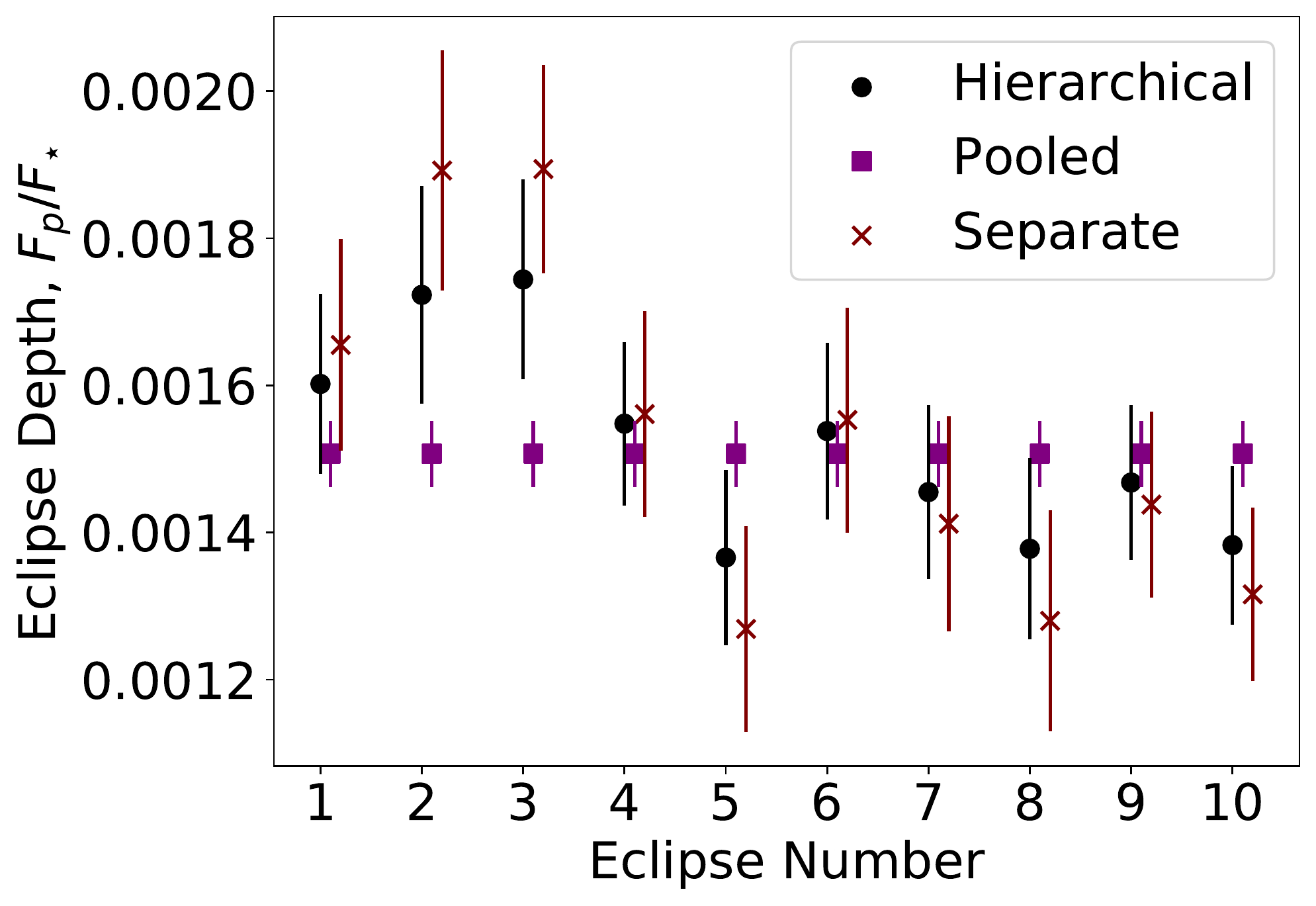}
\caption{Eclipse depths of XO-3b for the three different models. The pooled model has the lowest eclipse depth uncertainty, but is not the best model according to the WAIC. The hierarchical model is the best model, and yields eclipses that are closer together, with lower uncertainties on the individual eclipse depths than the separate model (15\% smaller on average). The hierarchical model represents a compromise between the overfit separate model and the underfit pooled model.
\label{XO-3b_moneyplot}}
\end{figure}

\begin{figure*}
\includegraphics[height=0.6\textheight]{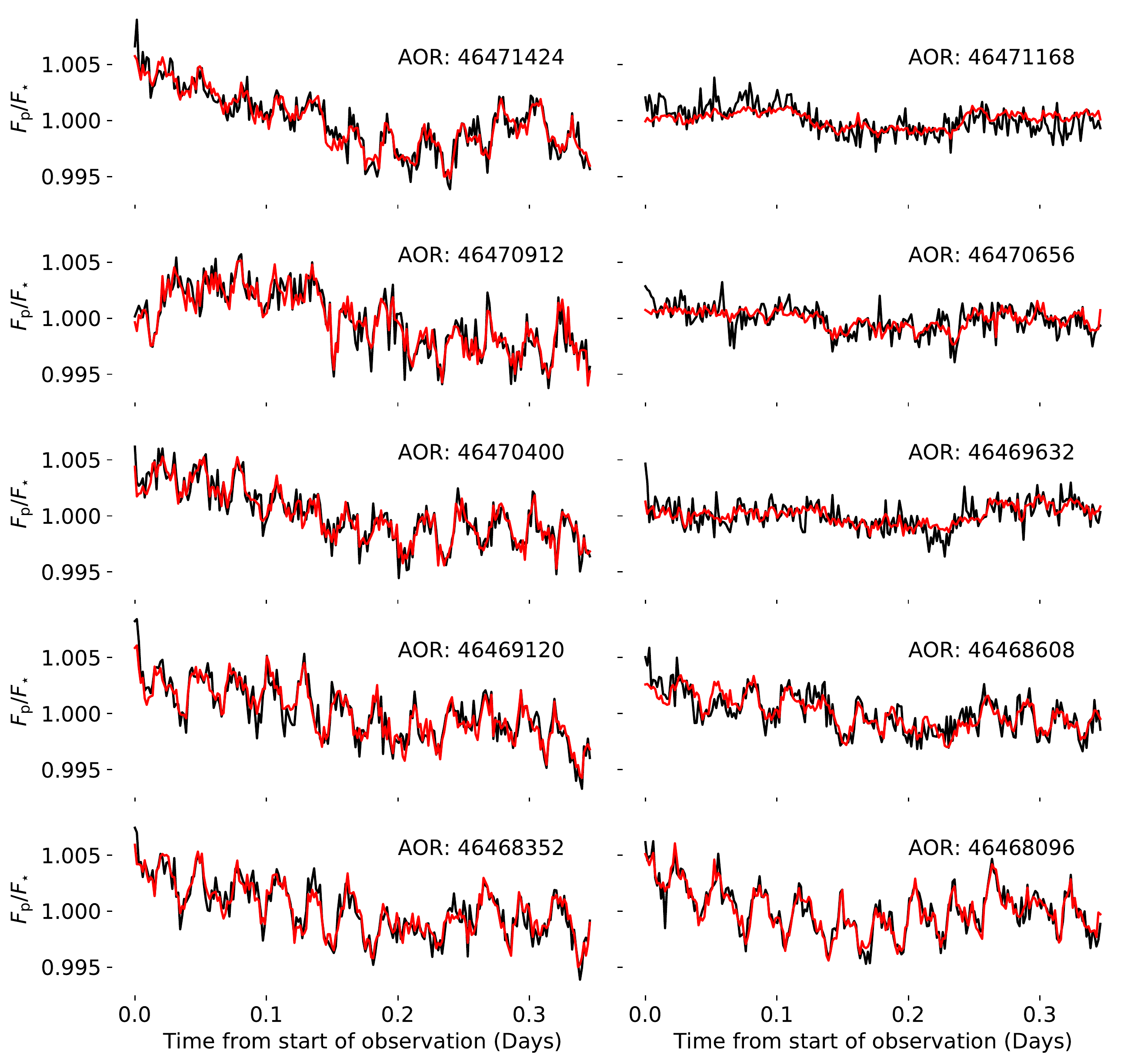}

\caption{The full fit to each of the ten XO-3b eclipses, using the hierarchical model for the eclipse depths and a pooled Gaussian process for the detector systematics.
\label{fig:XO-3b_eclipses}}
\end{figure*}

\begin{figure*}

\includegraphics[height=0.6\textheight]{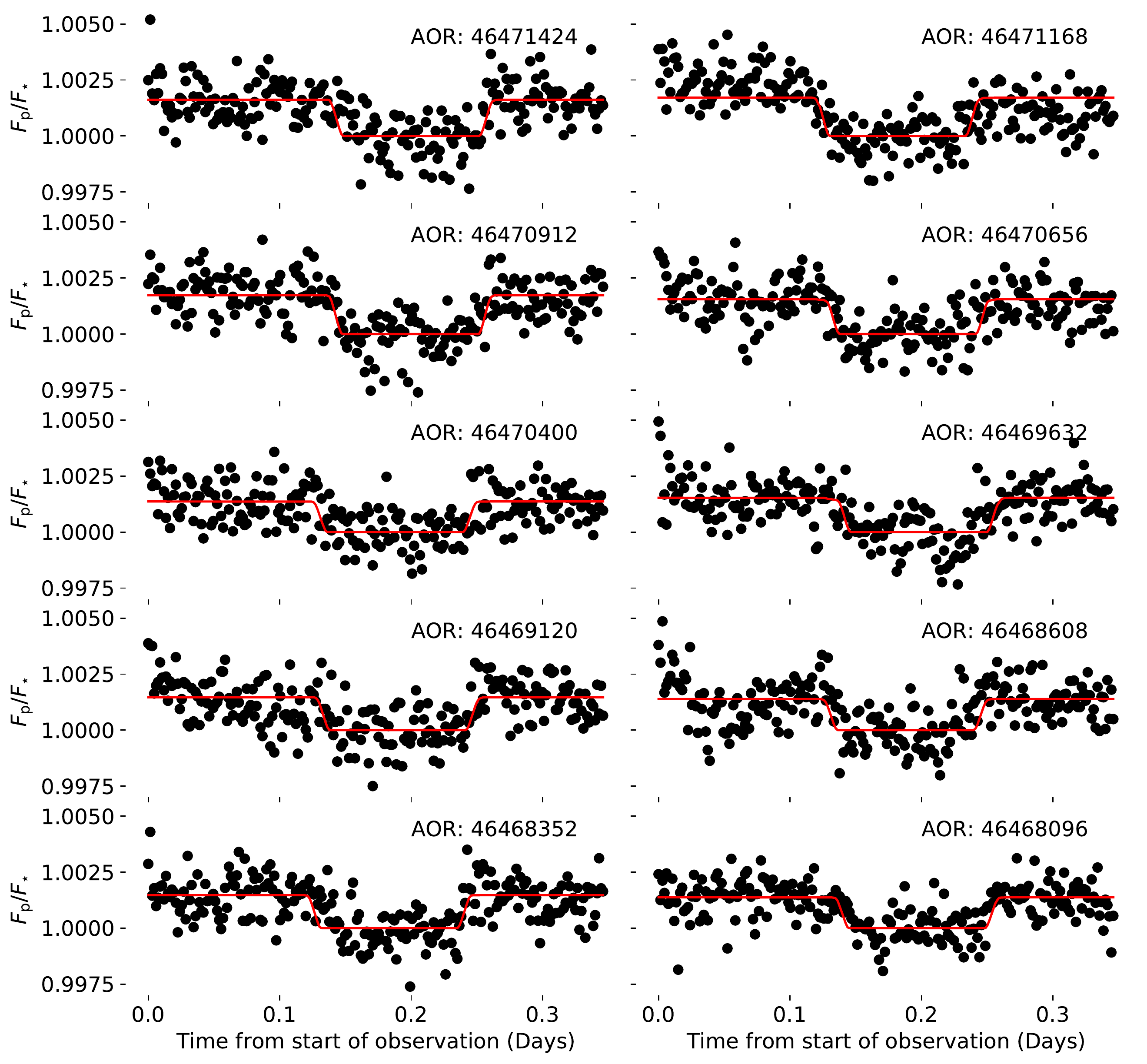}
\caption{The ten XO-3b eclipses with the detector systematics removed using a Gaussian process as a function of stellar centroid location. The eclipse signal is clearly visible in the corrected raw data (represented by the black dots), and the best fit eclipse signal for each is shown as the red line.
\label{fig:XO-3b_eclipses_signal}}
\end{figure*}

We fit the ten 4.5~$\mu$m eclipses of XO-3b with three models: a model where each eclipse observation had its own, separate eclipse depth parameter, one where we used a single, pooled eclipse depth for all the observations, and a hierarchical model. To fit each model, we used 2000 tuning steps to initialize four HMC chains, and obtained 1000 samples for each chain. After sampling, we confirmed that the Gelman-Rubin statistic was close to 1 for all parameter values, and that there were no divergences. The eclipse depths from each model are shown graphically in Figure~\ref{XO-3b_moneyplot} and tabulated in Table~\ref{EclipseDepths}. The best-fit model for each eclipse observation is shown in Figure~\ref{fig:XO-3b_eclipses}. Figure~\ref{fig:XO-3b_eclipses_signal} shows the best-fit eclipse signal after removing the detector systematics.   

We also computed the $\Delta$WAIC values compared to the best fit model, shown in Table~\ref{table:WAIC}. The hierarchical model had a significantly lower WAIC than the separate model, and a marginally lower WAIC than the single eclipse depth (pooled) model. This suggests that the eclipse depths are indeed different between observations, but are more similar than if we had used a separate eclipse parameter to describe each. This also hints at some variability from observation to observation. Typical hot Jupiters are predicted to show some epoch-to-epoch variability \citep{Komacek2020}, which is another reason to adopt the hierarchical model over the pooled one.
\begin{table}
	\centering
	\caption{Best-fit eclipse depths from each model of the ten \emph{Spitzer} IRAC channel 2 eclipses of XO-3b. The parameters $\mu$ and $\sigma$ are the hierarchical mean and standard deviation for the suite of eclipse depths. With the pooled model, we are implicitly assuming $\sigma = 0$, while for the separate model we are implicitly assuming $\sigma = \infty$. The hierarchical model fits for the amount of pooling, and is preferred over both the completely pooled and separate models. \label{EclipseDepths}}
	\label{tab:example_table}
	\begin{tabular}{cccc} 
	    Eclipse Number & Hierarchical (ppm) & Pooled (ppm) & Separate (ppm)\\
	    \hline
        1 & 1602$\pm$122 & 1507$\pm$45 & 1655$\pm$144 \\
        2 & 1723$\pm$148 & 1507$\pm$45 & 1892$\pm$163 \\
        3 & 1744$\pm$136 & 1507$\pm$45 & 1894$\pm$142 \\
        4 & 1548$\pm$111 & 1507$\pm$45 & 1561$\pm$140 \\
        5 & 1366$\pm$119 & 1507$\pm$45 & 1269$\pm$140 \\
        6 & 1538$\pm$120 & 1507$\pm$45 & 1553$\pm$153 \\
        7 & 1455$\pm$118 & 1507$\pm$45 & 1412$\pm$146 \\
        8 & 1378$\pm$123 & 1507$\pm$45 & 1280$\pm$150 \\
        9 & 1468$\pm$105 & 1507$\pm$45 & 1438$\pm$126 \\
        10 & 1383$\pm$108 & 1507$\pm$45 & 1316$\pm$118 \\
        \hline
        $\mu$ & 1520$\pm$81 & 1507$\pm$45 & 1527$\pm$219 \\
        $\sigma$ & 193$\pm$80 & 0 & $\infty$ \\ 
		\hline
	\end{tabular}
\end{table}

\begin{table}
	\centering
    \caption{Comparison of different information criteria for the three XO-3b eclipse depth models--- lower values are better. For the WAIC we also tabulate the standard error in the difference between each model. Using the BIC and AIC the pooled model is preferred. However, our model does not satisfy the assumptions necessitated by the BIC and AIC. The WAIC is more robust. According to the WAIC, the hierarchical model is the preferred model. 
    \label{table:WAIC}}
	\begin{tabular}{ccccc} 
	    Model &  $\Delta$ WAIC & $\sigma_{\Delta \rm WAIC}$ & $\Delta$AIC & $\Delta$BIC\\
	    \hline
        \textbf{Hierarchical} &  0.0 & 0.0 & 7.46 & 10.80  \\
        Pooled &  8.54 & 5.04 & 0 & 0 \\
        Separate &  65.66& 25.60 & 20.31  & 23.03\\
	\end{tabular}
\end{table}

The mean eclipse depth for the three models are consistent with one another within the uncertainties. In Figure~\ref{XO-3b_moneyplot}, we see the effects of shrinkage on the eclipse depths. Compared to the separate model, the hierarchical model yields smaller scatter across the suite of eclipse depths and higher precision on the individual eclipse depths. Additionally, the individual uncertainties on the fitted eclipse depths are smaller by 15\% on average in the hierarchical fits compared to the separate fits. The individual eclipse depth observations help constrain each other by shrinking the whole suite of eclipse depths towards the grand mean, but not as much as in the completely pooled model.

\newpage
\section{Hierarchical model for multiple planets}
\label{sec:multiplanet}
While hierarchical modelling is most obviously applicable for repeated measurements of the same planet, we can also extend it to secondary eclipse observations of multiple planets analyzed simultaneously. Specifically, we wanted to test the claim from both \citet{Garhart2020} and \citet{Baxter2020} that the 4.5~$\mu$m to 3.6~$\mu$m brightness temperature ratio increases with increasing stellar irradiation for hot Jupiters. We considered the observations from \citet{Garhart2020}, the largest dataset of uniformly reduced and analyzed hot Jupiter secondary eclipses.  

We expect that a hot Jupiter's dayside temperature, $T_{\rm d}$, is approximately proportional to its irradiation temperature, $T_{0}=T_{\rm eff}\sqrt{R_{\star}/a}$. We built a hierarchical model by including this intuition in our hyperprior, and making the hierarchical mean a function of irradiation temperature:
\begin{equation*}
    \mu_{\rm d} =m \left(T_{0}-\langle T_{0}\rangle \right) +b,
\end{equation*}
where $\langle T_{0}\rangle$ is the average irradiation temperature for the ensemble of planets. In other words, our hierarchical mean is now a line described by a slope and standard deviation in the $T_{\rm d}$ vs $T_{0}$ plane. We represent the scatter about this line using the hyperparameter $\sigma_{d}$. The prior on the dayside brightness temperature for a given planet is then $T_{\rm d,p} \sim \mathcal{N}(\mu_{\rm d}, \sigma_{d})$. 

For this hierarchical model, the hierarchical mean itself depends on two hyperparameters, the slope and intercept of the line, which we fit for simultaneously with the suite of dayside brightness temperatures. We used the following weakly informative priors for the hyperparameters: $m \sim \mathcal{N}(1, 0.5)$, $b \sim \mathcal{N}(2200K, 500K)$, $\sigma_{d} \sim$ Half-$\mathcal{N}(500K)$. 

We included the planets with measurements at both 4.5~$\mu$m and 3.6~$\mu$m, which gave a total of 33 planets. Fitting 66 eclipse observations simultaneously with a two-dimensional Gaussian process is computationally intractable using the hardware we had available, so we took the published measurements at face value rather than refit them. The reported eclipse depths and uncertainties are correct but were not properly propagated when converting to brightness temperature uncertainties (D.\ Deming, private communication), so we kept the sample of eclipses and eclipse depths measured by \citet{Garhart2020} but used the brightness temperatures and uncertainties calculated by \citet{Baxter2020}. Comparing the two datasets, \citet{Garhart2020} overestimated the brightness temperature uncertainties by about a factor of two for each planet.

We again used PyMC3, and first fit a non-hierarchical model as our baseline, using separate $T_{d,p}$ parameters for each eclipse. As expected, this model just reproduces the published dayside temperatures and uncertainties. This also acts as a confidence check that our priors are not biasing the fitted parameters. 

\begin{figure}
\includegraphics[width=\columnwidth]{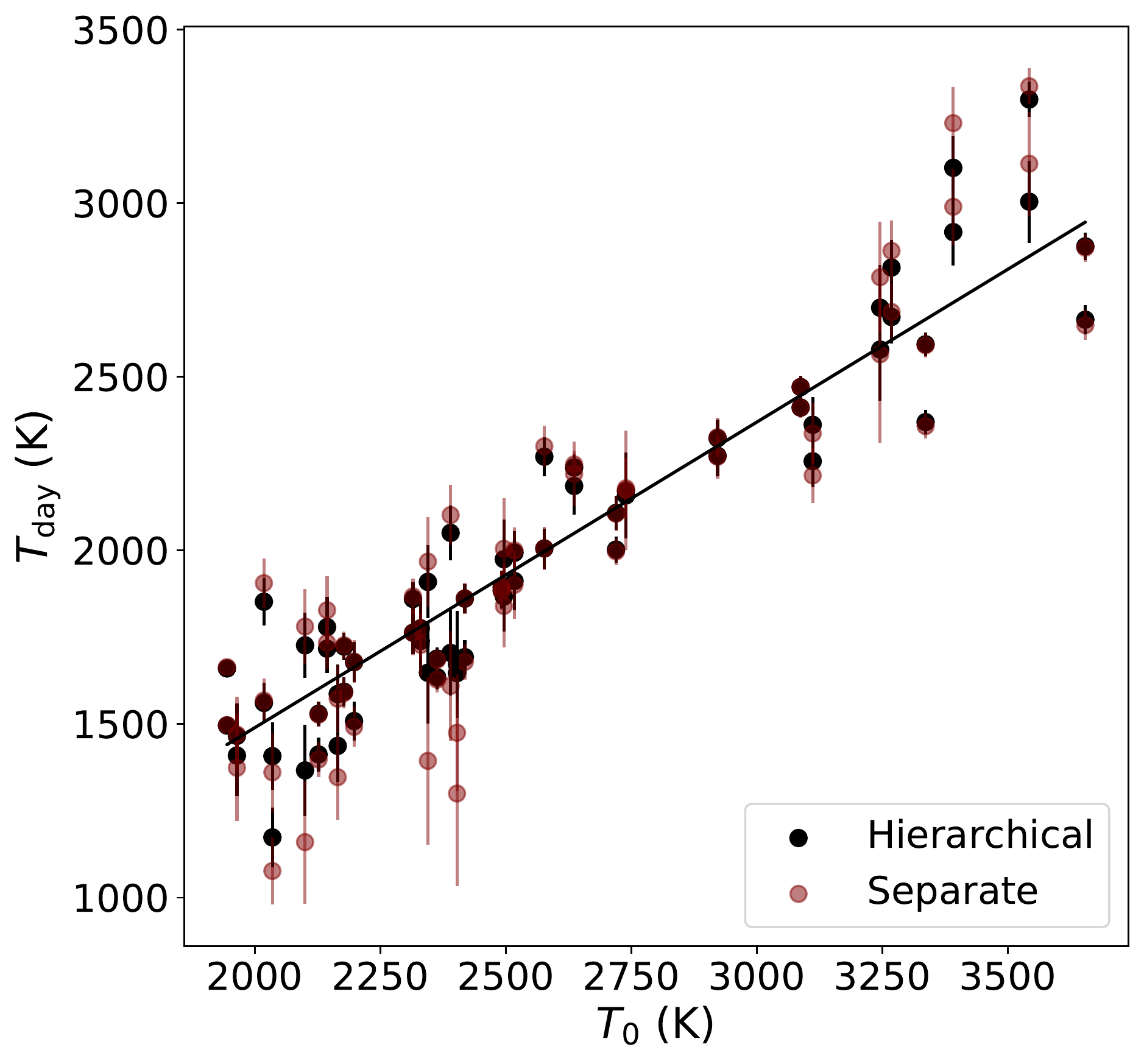}
\caption{Dayside brightness temperatures, at 3.6~$\mu$m and 4.5~$\mu$m, for the planets analyzed by \citet{Garhart2020}, refit with a wavelength independent hierarchical model. The red dots are the published values and the black line is the best fit trend line. The effects of Bayesian shrinkage are evident: the measurements are clustered closer to the line, and the uncertainties on the measurements are reduced.
\label{Garhart_moneyplot}}
\end{figure}

Since there are measurements at two different wavelengths, we fit two different versions of the hierarchical model. In the wavelength dependent model, we allowed the dayside brightness temperature distributions to be different between the two wavelengths, fitting one set of hierarchical parameters for the 4.5~$\mu$m measurements, and another set for the 3.6~$\mu$m measurements. In the wavelength independent model, we used a common distribution for all the measurements, and thus one set of hierarchical parameters. We tabulate the refit brightness temperatures in Table~\ref{table:Garhart2020_Fit} and plot them in Figure~\ref{Garhart_moneyplot}.
\begin{table*}
	\centering
    \caption{Results from fitting a wavelength independent hierarchical model to the \citet{Garhart2020} eclipse dataset. The columns $T_{\rm d, ch2}$ and $T_{\rm d, ch1}$ list the refit dayside brightness temperatures at 4.5~$\mu$m 3.6~$\mu$m. 
\label{table:Garhart2020_Fit}}
\begin{tabular}{llll}
Planet     & $T_{0} (K)$      & $T_{\rm d, ch2}$ (K)     & $T_{\rm d, ch1}$ (K)        \\
\hline
HAT-P-13 b  & 2331$\pm$75  & 1739$\pm$81  & 1776$\pm$79  \\
HAT-P-30 b  & 2315$\pm$61  & 1763$\pm$61  & 1860$\pm$49  \\
HAT-P-33 b  & 2517$\pm$48  & 1912$\pm$85  & 1993$\pm$63  \\
HAT-P-40 b  & 2496$\pm$93  & 1867$\pm$100 & 1975$\pm$113 \\
HAT-P-41 b  & 2739$\pm$62  & 2171$\pm$77  & 2158$\pm$124 \\
KELT-2 A b  & 2418$\pm$44  & 1693$\pm$49  & 1861$\pm$42  \\
KELT-3 b    & 2577$\pm$62  & 2006$\pm$58  & 2270$\pm$57  \\
Qatar-1 b   & 1964$\pm$61  & 1466$\pm$93  & 1409$\pm$117 \\
WASP-100 b  & 3111$\pm$242 & 2362$\pm$80  & 2257$\pm$74  \\
WASP-101 b  & 2198$\pm$57  & 1509$\pm$56  & 1678$\pm$58  \\
WASP-103 b  & 3543$\pm$110 & 3299$\pm$51  & 3005$\pm$119 \\
WASP-104 b  & 2144$\pm$61  & 1779$\pm$88  & 1717$\pm$70  \\
WASP-12 b   & 3654$\pm$129 & 2665$\pm$42  & 2876$\pm$40  \\
WASP-121 b  & 3336$\pm$86  & 2594$\pm$34  & 2370$\pm$35  \\
WASP-131 b  & 2035$\pm$51  & 1174$\pm$86  & 1408$\pm$98  \\
WASP-14 b   & 2636$\pm$85  & 2186$\pm$83  & 2239$\pm$38  \\
WASP-18 b   & 3391$\pm$103 & 3102$\pm$92  & 2917$\pm$96  \\
WASP-19 b   & 2922$\pm$65  & 2273$\pm$59  & 2323$\pm$53  \\
WASP-36 b   & 2403$\pm$64  & 1647$\pm$125 & 1672$\pm$154 \\
WASP-43 b   & 1945$\pm$112 & 1496$\pm$24  & 1660$\pm$24  \\
WASP-46 b   & 2345$\pm$78  & 1910$\pm$105 & 1648$\pm$146 \\
WASP-62 b   & 2018$\pm$49  & 1561$\pm$58  & 1852$\pm$68  \\
WASP-63 b   & 2165$\pm$64  & 1437$\pm$104 & 1586$\pm$85  \\
WASP-64 b   & 2390$\pm$74  & 1705$\pm$122 & 2051$\pm$79  \\
WASP-65 b   & 2100$\pm$83  & 1367$\pm$131 & 1727$\pm$94  \\
WASP-74 b   & 2720$\pm$75  & 2108$\pm$49  & 2003$\pm$38  \\
WASP-76 b   & 3087$\pm$66  & 2471$\pm$32  & 2412$\pm$28  \\
WASP-77 A b & 2363$\pm$44  & 1635$\pm$36  & 1689$\pm$31  \\
WASP-78 b   & 3246$\pm$124 & 2579$\pm$148 & 2699$\pm$123 \\
WASP-79 b   & 2492$\pm$75  & 1885$\pm$52  & 1895$\pm$47  \\
WASP-87 b   & 3268$\pm$96  & 2815$\pm$79  & 2673$\pm$76  \\
WASP-94 A b & 2127$\pm$109 & 1412$\pm$49  & 1530$\pm$35  \\
WASP-97 b   & 2178$\pm$59  & 1593$\pm$43  & 1723$\pm$39
\end{tabular}
\end{table*}

We show the difference in WAIC values for the three models in Table~\ref{table:WAIC_Garhart2020}. The wavelength independent model did slightly better than the wavelength dependent model ($\Delta$ WAIC = 0.54), however the uncertainty on that difference is 1.35, meaning the models make equally good predictions. In the wavelength independent model, the dayside brightness temperatures for both channels follow the same slope, or equivalently, the ratio of the slopes for each channel is equal to one. This means we are not detecting---nor ruling out---the trend of increasing brightness temperature ratio versus stellar irradiation reported by \citet{Garhart2020} and \citet{Baxter2020}. The best-fit hierarchical parameters from our wavelength independent model are $\mu_{d} = 1.24 \pm 0.06$, $b = 2003 \pm 24 K$, and $\sigma_{m} = 181 \pm 20 K$.

The difference in WAIC suggests that the non-hierarchical model makes equally good predictions compared to the wavelength independent hierarchical model (Table~\ref{table:WAIC_Garhart2020}), which is at odds with expectations that irradiation temperature should determine planetary dayside temperatures.

\begin{table}
	\centering
    \caption{WAIC scores for the three models used to fit the suite of eclipses from \citet{Garhart2020}--- lower values are better. We also tabulate the standard error of the difference in WAIC between each model. The three models make equally good predictions according to the WAIC scores.  
\label{table:WAIC_Garhart2020}}
	\begin{tabular}{ccc} 
	    Model &  $\Delta$ WAIC & $\sigma_{\Delta \rm WAIC}$ \\
	    \hline
        Separate  &  0.0 & 0.0  \\
        Wavelength Independent & 1.13 & 2.54\\
        Wavelength Dependent &  1.67 & 2.42 \\
	\end{tabular}
\end{table}

\begin{figure}
\includegraphics[width=\columnwidth]{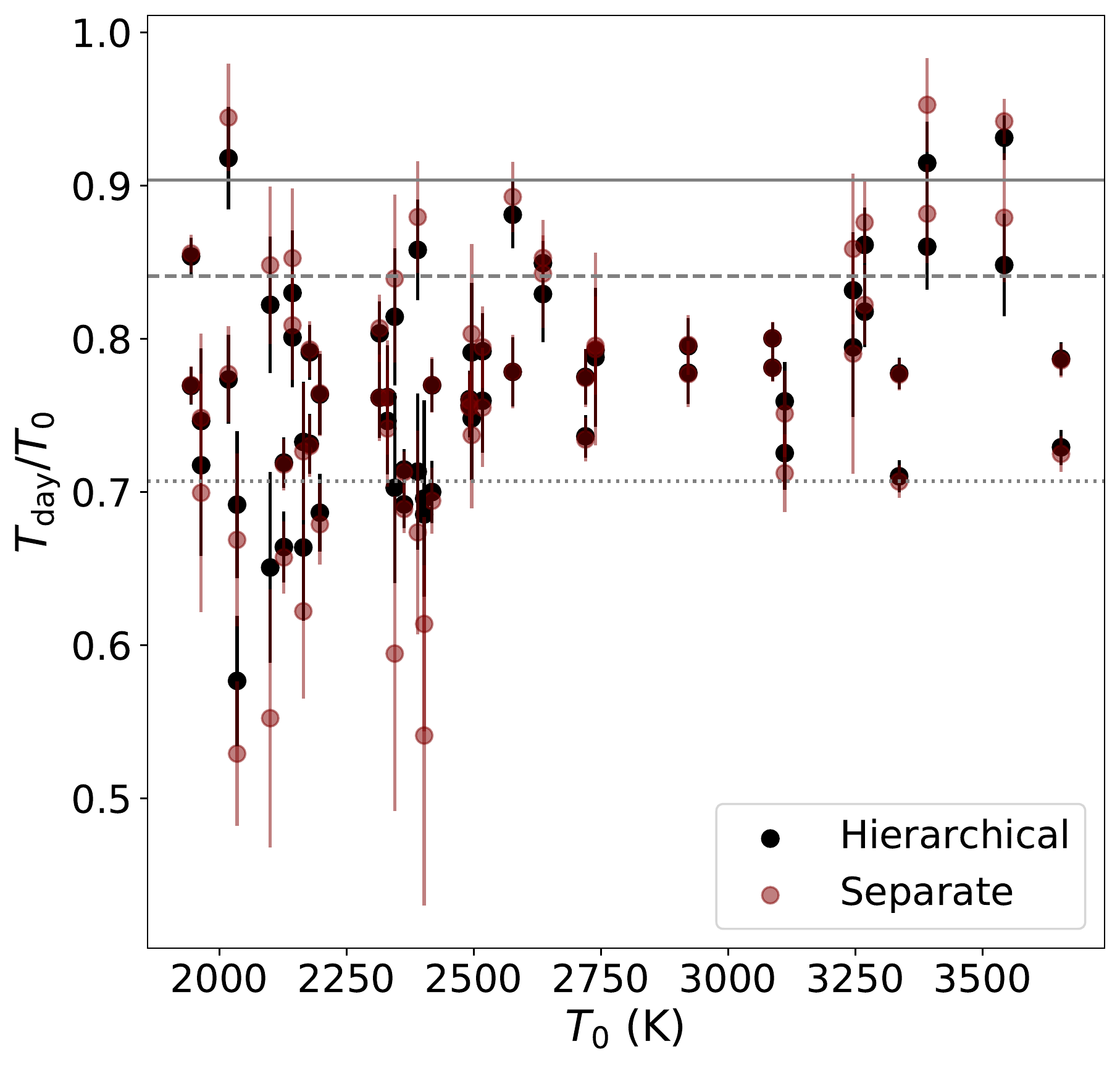}
\caption{Dayside brightness temperatures from Fig~\ref{Garhart_moneyplot}, scaled by irradiation temperature. The horizontal lines represent some theoretical limits on dayside temperature assuming a zero Bond albedo: the solid line assumes zero heat recirculation, the dashed line assumes a uniform dayside hemisphere but a temperature of zero on the nightside, and the dotted line assumes a uniform temperature at every location on the planet. 
\label{Garhart_scaled}}
\end{figure}

\section{Discussion and conclusions}
\subsection{Repeat observations of a single planet}
In our reanalysis of the ten secondary eclipses of XO-3b, we found that the hierarchical model was favoured over the two non-hierarchical models. This means that the measured eclipse depths are indeed different from epoch to epoch, yet clustered. The biggest difference compared to previous analyses is that we were able to empirically fit for the amount of epoch-to-epoch scatter favoured by the data, and doing this improves the precision on our measurements by 15\% on average, because of Bayesian shrinkage. Notably, we found that the hierarchical eclipse depth had a larger standard deviation than the individual measurements, suggesting that measuring just one eclipse depth could lead one to underestimate the true uncertainty compared to the hierarchical approach.  

Hierarchical models could improve measurements of other hot Jupiters and other types of planets. Hierarchical models could be used to robustly test the reported variability in the secondary eclipses of the super earth 55 Cancri e \citep{Demory2016,Tamburo2018}, or to fit the twelve \textit{Spitzer} eclipses of the recently discovered hot Saturn LTT 9779b \citep{Dragomir2020}. The published variability constraints for HD 189733b \citep{Agol2010} and HD 209458b \citep{Kilpatrick2020} could also be revisited with hierarchical models. 

Repeated phase curve observations could benefit from using hierarchical models. The hot Jupiter WASP-43b has one published \citep{Stevenson2017,Mendonca2018,Morello2019,May2020,Bell2021}, and two unpublished, \textit{Spitzer} phase curves at 4.5~$\mu$m. A hierarchical model could be used to better constrain the phase amplitudes and offsets of the three 4.5~$\mu$m phase curves by fitting them simultaneously.

To test whether we are seeing the effects of variability or detector systematics, the best approach is to compare planets with repeated observations in both \textit{Spitzer} channels. If certain types of planets have larger hierarchical eclipse standard deviations, the culprit could be time variability that is only exhibited by certain planets. Otherwise, if one \textit{Spitzer} channel tends show more variability regardless of planet, it suggests that detector systematics are at play. Spectroscopic observations will also be able to break the degeneracy between variability and detector systematics, as would simultaneous measurements with multiple instruments.

\subsection{Parallel analysis of multiple planets}
We showed that our hierarchical model of measurements from multiple planets yields smaller uncertainties on the individual eclipse depths, and tends to shrink the eclipse depths toward the trend line. We did not detect the trend of increasing brightness temperature ratio with increasing stellar irradiation reported by \cite{Garhart2020} and \cite{Baxter2020}, nor did we rule it out. The hierarchical models made predictions that were as good as the non-hierarchical model, when comparing the WAIC values. 

One possible explanation is that the uncertainties on the eclipse depths are underestimated due to detector systematics or astrophysical variability. This was first suggested by \citet{Hansen2014}, who concluded that the first generation of \emph{Spitzer} eclipse uncertainties may be underestimated by up to a factor of 3, probably due to inadequate treatment of detector systematics. Hot Jupiter infrared eclipse depths are generally assumed to be the same from epoch-to-epoch because most general circulation models produce stable circulation patterns \citep{Komacek2020}, but recent work using high-resolution GCMs predicts multiple equilibria in hot Jupiter atmospheres and transient planetary-scale storms \citep{Cho2021}. The consequence of such variability, much like detector systematics, is that measuring just a single eclipse for a planet in a given bandpass would lead one to underestimate the uncertainty.
Indeed, we found that for XO-3b, the hierarchical standard deviation was larger than the individual uncertainties by about a factor of 1.5--2. This suggests that if we had observed only one eclipse of XO-3b, we would have underestimated the eclipse depth uncertainty compared to the estimate from the hierarchical model.

In the context of a hierarchical model, small measurement uncertainties leave less leeway for Bayesian shrinkage. Indeed, repeating our analysis using the larger, albeit miscalculated, uncertainties from \citet{Garhart2020} showed a marked improvement when using the hierarchical model compared to the completely separate model (see the Appendix for the results of that analysis). 

Another explanation for the marginal performance of hierarchical models on the \citet{Garhart2020} ensemble of planets is that irradiation temperature is not the sole determinant of planetary dayside temperatures. It is becoming clear that secondary parameters like planetary mass, radius, and rotation rate play important roles in determining atmospheric circulation on hot Jupiters \citep{Keating2019,Bell2021}. Differences in these parameters could contribute additional planet-to-planet scatter.

In this work we used the largest subset of \textit{Spitzer} secondary eclipses that had been uniformly reduced and analyzed. One obvious extension of our work is to refit the detector systematics and astrophysical signals for all \textit{Spitzer} secondary eclipses using a uniform pipeline. We recommend using a hierarchical model for the dayside brightness temperatures and placing a second level of hierarchy on the planets with repeated eclipses. This would take a prohibitively long time using a two dimensional Gaussian process and conventional hardware like we did for XO-3b, but it could potentially be done using high-performance or GPU computing. Alternatively, such a fit could be done using an easier-to-compute detector model like Pixel Level Decorrelation \citep{Deming2015, Garhart2020}, especially with PyMC3.

In this work we have shown that hierarchical models are useful when analyzing repeated measurements from a single target, or when doing comparative exoplanetology of many targets. Next generation telescopes like \textit{James Webb} and \emph{Ariel} will make repeated measurements of certain targets, and will both carry out photometric and spectroscopic transit, eclipse, and phase curve surveys for a variety of targets \citep{Bean2018,Tinetti2018,Charnay2021}. This will allow for atmospheric characterization of potentially thousands of more exoplanets, from Earth-like planets to ultra-hot Jupiters, and we recommend that these comparative surveys incorporate hierarchical modelling to make measurements and predictions that are as robust as possible.

\section*{Acknowledgements}
This project was conceived at the ``Multi-dimensional characterization of distant worlds: spectral retrieval and spatial mapping'' workshop hosted by the Michigan Institute for Research in Astrophysics and spearheaded by Emily Rauscher.  We are particularly grateful to David van Dyk for a pedagogical introduction to Bayesian shrinkage.  We acknowledge support from the McGill Space Institute
and l'Institut de recherche sur les exoplan\`etes. We have made use of open-source software provided by the Python, Astropy, SciPy, Matplotlib, and PyMC3 communities.

\section*{Data Availability}

The reduced photometry for the ten archival secondary eclipse observations of XO-3b are freely available at \url{https://irachpp.spitzer.caltech.edu/page/data-challenge-2015}. The code used in the XO-3b reanalysis, and a Jupyter notebook showing the reanalysis of the \citet{Garhart2020} eclipses can both be found at \url{https://github.com/dylanskeating/HARMONiE}.



\bibliographystyle{mnras}
\bibliography{mnras_template} 




\appendix
\label{appendix}
\section{Analysis using inflated uncertainties}
We also considered the brightness temperatures and uncertainties reported by \citet{Garhart2020}. Their eclipse depths, eclipse depth uncertainties, and brightness temperatures are correct, but the brightness temperature uncertainties were derived by taking the relative uncertainty in eclipse depth and using that to calculate the uncertainty in brightness temperature (D. Deming, private communication). In their reanalysis, \citet{Baxter2020} used the eclipse depths and uncertainties reported by \citet{Garhart2020} but fully propagated those uncertainties through the Planck function, which is non-linear, to derive uncertainties on the brightness temperatures. Comparing the uncertainties reported in both works, the uncertainties of \citet{Garhart2020} are roughly twice as big as those reported by \citet{Baxter2020}. 

To see how our conclusions would change had we used the artificially inflated uncertainties, we refit the multi-planet hierarchical model from Section~\ref{sec:multiplanet}. According the $\Delta$WAIC scores (Table~\ref{table:WAIC_Garhart2020_app}), the wavelength independent hierarchical model makes much better predictions than the non-hierarchical model, and marginally better predictions than the wavelength dependent model. Again, we do not detect the trend of increasing 4.5~$\mu$m to 3.6~$\mu$m brightness temperature ratio, nor do we rule it out.

\begin{figure}
\includegraphics[width=\columnwidth]{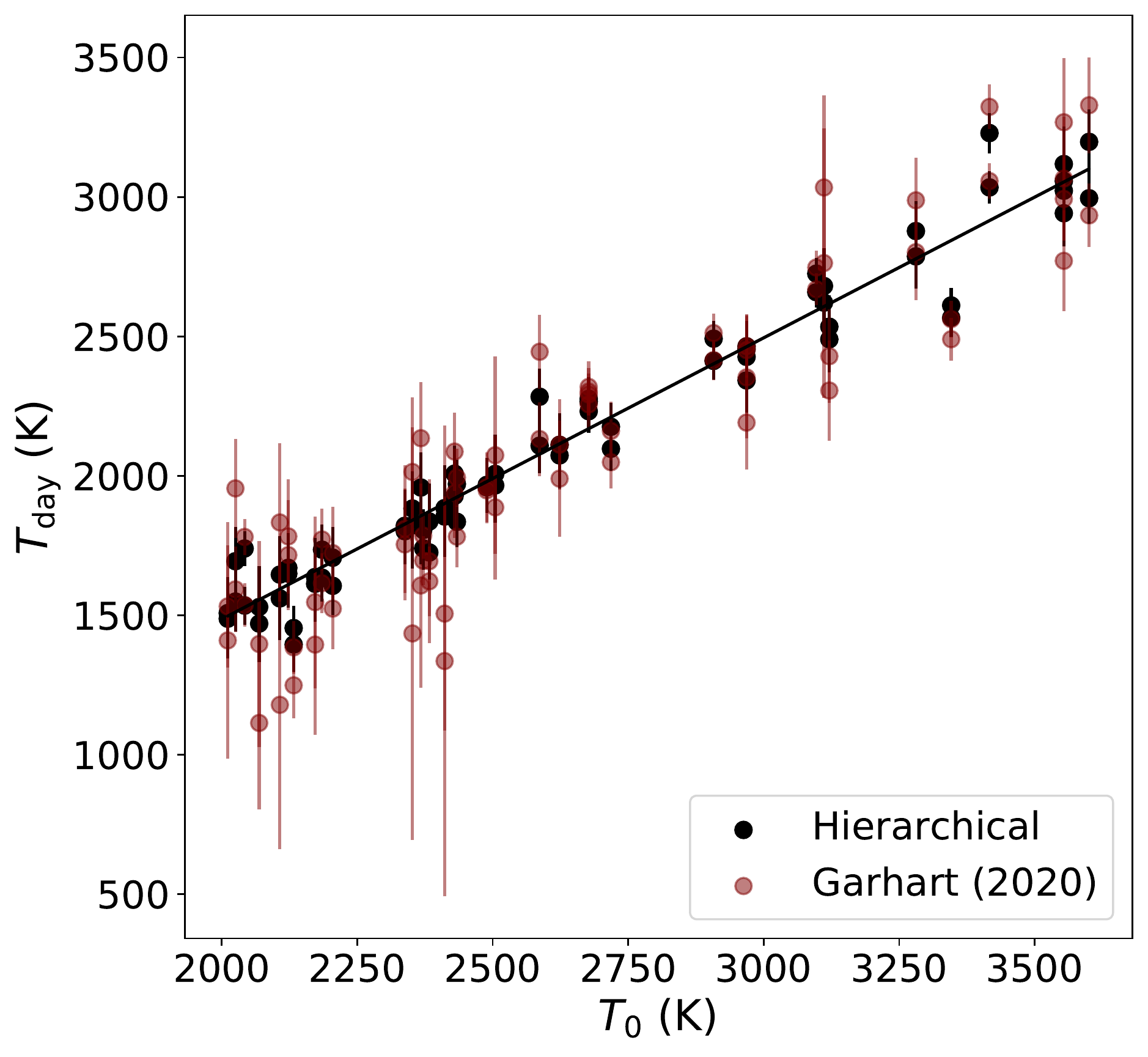}
\caption{Dayside brightness temperatures refit with a wavelength independent hierarchical model. We used the inflated uncertainties from \citep{Garhart2020}. The red dots are the published values and the black line is the best fit trend line. The effects of Bayesian shrinkage are evident: the measurements are clustered closer to the line, and the uncertainties on the measurements are reduced.
\label{Garhart_moneyplot_app}}
\end{figure}

\begin{figure}
\includegraphics[width=\columnwidth]{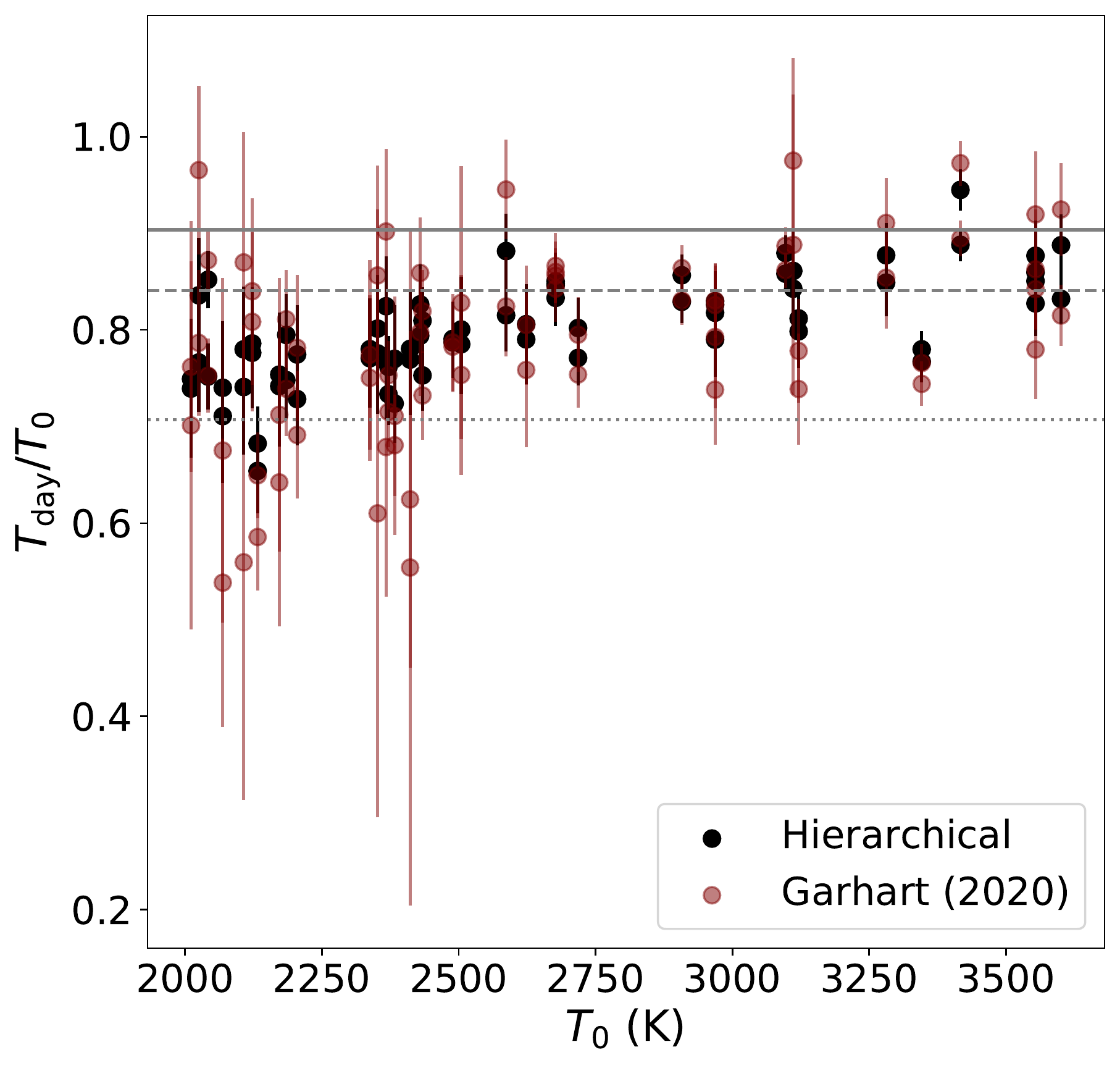}
\caption{Dayside brightness temperatures from Fig~\ref{Garhart_moneyplot_app}, this time scaled by irradiation temperature. The horizontal lines represent some theoretical limits on dayside temperature assuming a zero Bond albedo: the solid line assumes zero heat recirculation, the dashed line assumes a uniform dayside hemisphere but a temperature of zero on the nightside, and the dotted line assumes a uniform temperature at every location on the planet. 
\label{Garhart_scaled_app}}
\end{figure}

The refit brightness temperatures are shown in Figure~\ref{Garhart_moneyplot_app} and Figure~\ref{Garhart_scaled_app}. When measurement uncertainties are higher, Bayesian shrinkage is more dramatic. 

\begin{table}
	\centering
    \caption{WAIC scores for the three models, using the artificially inflated brightness temperature uncertainties from \citep{Garhart2020}. We also tabulate the standard error of the difference in WAIC between each model. The two hierarchical models both make equally good predictions, and significantly outperform the non-hierarchical model.  
\label{table:WAIC_Garhart2020_app}}
	\begin{tabular}{ccc} 
	    Model &  $\Delta$ WAIC & $\sigma_{\Delta \rm WAIC}$ \\
	    \hline
        Wavelength Dependent  &  0.0 & 0.0  \\
        Wavelength Independent &  0.25 & 1.8\\
        Separate &  11.09& 3.61 \\
	\end{tabular}
\end{table}

\bsp	
\label{lastpage}
\end{document}